# Aberration associated with the reflection of light on a moving mirror

**Bernhard Rothenstein and Ioan Damian**

Politehnnica University of Timisoara, Department of Physics,
Timisoara, Romania
**bernhard_rothenstein@yahoo.com, ijdamian@yahoo.com**

**Abstract**. Imposing start from the beginning that the incidence and the reflection of a ray t on an arbitrarily orientated mirror take place at the same point in space and at the same zero time in all involved reference frames in relative motion, we decompose the reflection in two aberration of light effects, of the incident and of the reflected rays respectively.

### 1. Introduction

The reflection of a plane-polarized wave on a moving mirror is associated with the aberration of light effect.[1,2,3,4,5,6]. It is known that the formulas that account for aberration can be derived from the constant light speed postulate, without using the Lorentz-Einstein transformations[6]. Most of the papers covering this topics deal with the case of a vertical mirror the plane of which is normal to the direction in which it moves.

The purpose of our paper is to consider the general case of an arbitrary orientated mirror. Figure 1 shows a plane-polarized light wave incident on a plane mirror. The mirror is at rest in the K'(X'O'Y') inertial reference frame. One of the incidence points coincides with the origin O' and takes place at a time t'=0, generating the event $O'(x'=0, y'=0, t'=0)$ that is characterized by the same space-time coordinates in all inertial reference frames in relative uniform motion. In the rest frame of the mirror, the angles of incidence and reflection are equal to each other and equal to $\alpha$.

The problem is to find a relationship between the angle of incidence *i* and the angle of reflection *r* both measured in the K(XOY) reference frame, relative to which K' moves with constant velocity *V* in the positive direction of the common OX(O'X') axes. The axes of the two reference frames are parallel to each other and at the origin of time in the two frames (*t=t'=0*) the origins O and O' are located at the same point in space. The condition imposed above makes that in both reference frames incidence and reflection take place at the same point in space and at the same time.

Even if it is considered that the use of the Lorentz-Einstein transformations when we derive for relativistic phenomena is straightforward but not very transparent from the point of view of physics[7] we will use them, carefully defining the space-time coordinates of the involved events.

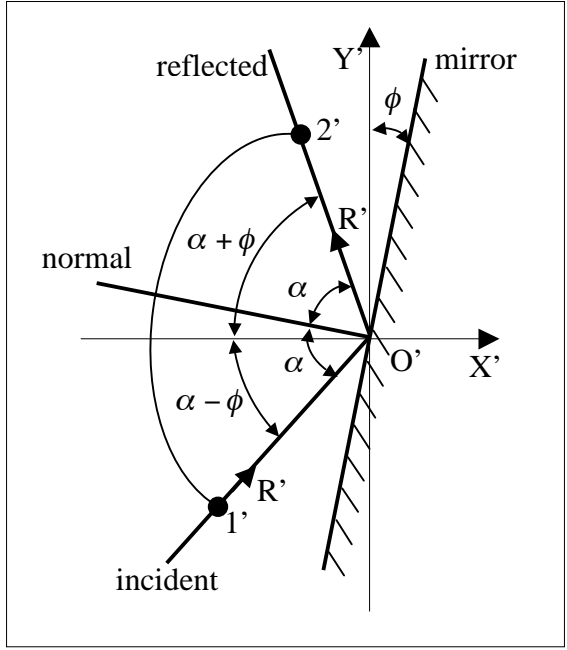

Fig. 1. Reflection of a light ray on an arbitrarily orientated plane mirror as detected from its rest frame.

## 2. Aberration of light associated with reflection of light on a moving mirror

Consider the event **1'**$[x_1' = -R'\cos(\alpha - \phi), y_1' = -R'\sin(\alpha - \phi), t_1' = -\dfrac{R'}{c}]$ taking place on the incident ray and **2'**$[x_2' = -R'\cos(\alpha + \phi), y_2' = R'\sin(\alpha + \phi), t_2' = \dfrac{R'}{c}]$ taking place on the reflected ray, both located at the same distance $R'$ from the incidence point in K'. In accordance with the Lorentz-Einstein transformations event **1'** is characterized in K by the space coordinates

$$x_1 = -\gamma R'[\cos(\alpha - \phi) + \beta] \tag{1}$$
$$y_1 = -R'\sin(\alpha - \phi). \tag{2}$$

Therefore the incidence angle $i$ is given by

$$\tan i = \frac{y_1}{x_1} = \gamma^{-1} \frac{\sin(\alpha - \phi)}{\cos(\alpha - \phi) + \beta} \tag{3}$$

where we have used the accepted relativistic notations $\beta = \dfrac{V}{c}$ and $\gamma = \dfrac{1}{\sqrt{1-\beta^2}}$. Event **2'** is characterized in K by the space coordinates

$$x_2 = \gamma R'[-\cos(\alpha + \phi) + \beta] \tag{4}$$
$$y_2 = R'\sin(\alpha + \phi). \tag{5}$$

Combing (4) and (5) for the reflection angle $r$ measured in K we obtain

$$\tan r = \frac{y_2}{x_2} = -\gamma^{-1} \frac{\sin(\alpha + \phi)}{\cos(\alpha + \phi) - \beta}. \tag{6}$$

whereas from (3) we obtain the reflection law in the K frame

$$\frac{\tan r}{\tan i} = -\frac{\sin(\alpha + \phi)}{\sin(\alpha - \phi)} \frac{\cos(\alpha - \phi) + \beta}{\cos(\alpha + \phi) - \beta}. \tag{7}$$

Equation (7) is a genuine transformation equation, containing in its left hand side only physical quantities measured in K whereas in its right hand side only physical quantities measured in K'. From a practical point of view it enables observers from K' to describe the reflection in K as a function of physical quantities measured in theirs own reference frame.

We consider two particular cases. First the case of the vertical mirror ($\phi = 0$) in which (7) becomes

$$\left(\frac{\tan r}{\tan i}\right)_{\phi=0} = -\frac{\cos\alpha + \beta}{\cos\alpha - \beta}. \tag{8}$$

In the case of a horizontal mirror ($\phi = \frac{\pi}{2}$) equation (7) becomes

$$\left(\frac{\tan r}{\tan i}\right)_{\phi=\frac{\pi}{2}} = -1 \tag{9}$$

or

$$r = -i \tag{10}$$

therefore, reflection law, as a true law of nature, holds in all inertial reference frames. In accordance with the principle of relativity it should hold for all values of $\phi$ mainly because as detected from K the mirror rotates in such a way that the reflection law is preserved. As customary in special relativity its formulas depend on the direction of the relative velocity and on the direction in which the rays of light propagate. Changing the direction of relative motion ($V \to -V$) or changing the order incidence-reflection ($c \to -c$) but not both of them at once (7) becomes

$$\frac{\tan r}{\tan i} = -\frac{\sin(\alpha + \phi)}{\sin(\alpha - \phi)}\left[\frac{\cos(\alpha - \phi) - \beta}{\cos(\alpha + \phi) + \beta}\right]. \tag{11}$$

In order to illustrate the results obtained above we consider (3) and (6) in the case of a vertical mirror

$$i = \arctan\left[\gamma^{-1}\frac{\sin\alpha}{\cos\alpha + \beta}\right] \tag{12}$$

$$r = \arctan\left[-\gamma^{-1}\frac{\sin\alpha}{\cos\alpha - \beta}\right] \tag{13}$$

Figure 2 presents a plot of *i* and *r* as a function of $\alpha$ for $\beta = 0.6$. It enables us to find out the relationship between *i* and *r* for the same value of $\alpha$ and to have a transparent look at the way in which motion influences reflection on a plane mirror. Figure 3 presents a plot of r as a function of i with $\alpha$ and $\beta$ as a parameter.

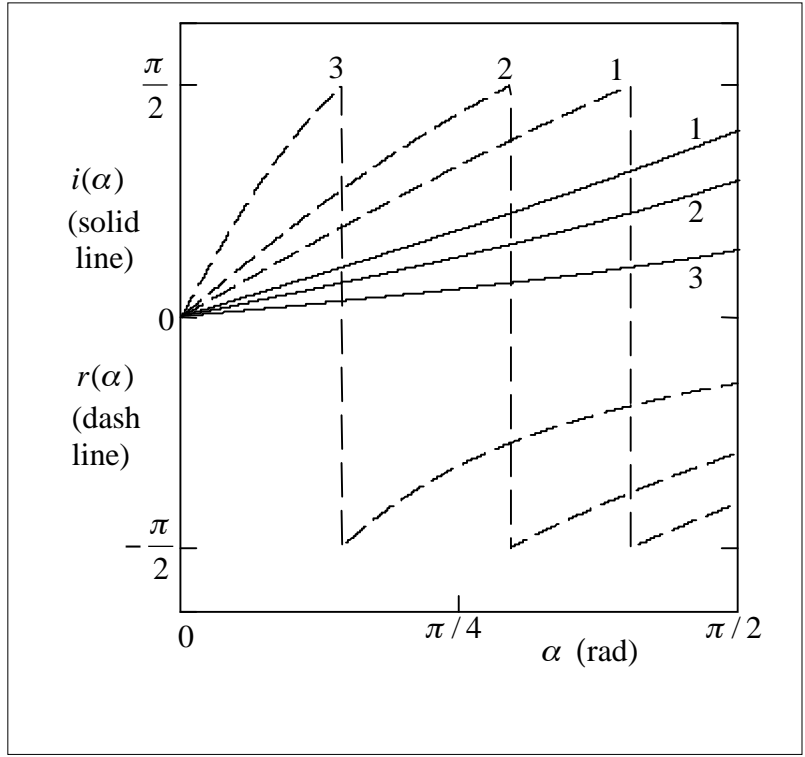

Figure 2. A plot of the incidence angle $i$ and of the reflection angle $r$ as a function of the angle $\alpha$ that characterizes the reflection in the mirror's rest frame, for different values of the relative velocity $V$, ($\beta = Vc^{-1}$).

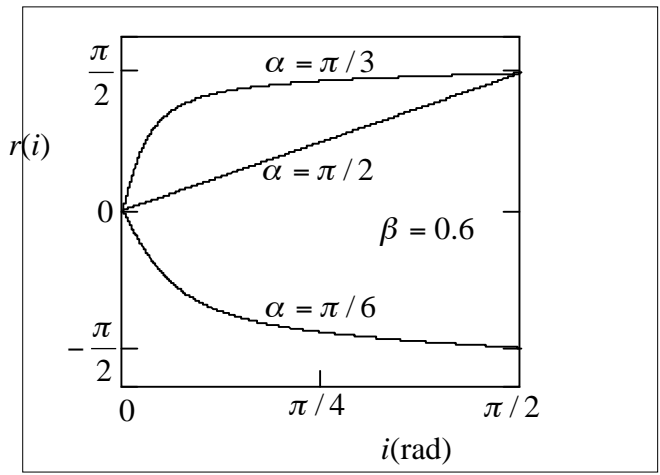

Figure 3. A plot of the reflection angle $r$ as a function of the incidence angle $i$ with $\alpha$ and $\beta$ as parameters.

### 3. Does the reflection law work in all inertial reference frames?

We have shown so far that the reflection law works in the case of a horizontal mirror in all inertial reference frames in relative motion. Consider a ray that propagates parallel to the surface of the mirror $\left(\alpha = \dfrac{\pi}{2}\right)$ when detected from K', where it is not deflected by the mirror. When detected from K that ray propagates along a direction that makes an angle $\phi$ with the positive direction of the common OX(O'X') axes, given by

$$\tan\phi = \frac{\gamma^{-1}}{\beta}. \tag{14}$$

The principle of relativity requires that when detected from K the mirror remains a plane one and that the incidence and the reflection characterized in K' by the angle $\alpha$ should look as shown in Figure 1. If the reflection law works in K we should have

$$r - i = \pi - \phi \tag{15}$$

Expressing (15) as a function of $r$, $i$ and $\phi$, it becomes an identity, which is $\alpha$ independent the reflection law working in all inertial reference frames in relative motion.

### 4. Conclusions

We have shown that imposing a certain initial condition, we can reduce the reflection of light on a plane mirror to the aberration of two rays that propagate in a given inertial reference frame symmetrically with respect to the common axes but in opposite direction.